\documentclass[fleqn,usenatbib]{mnras}

\usepackage{newtxtext,newtxmath}

\usepackage[T1]{fontenc}

\DeclareRobustCommand{\VAN}[3]{#2}
\let\VANthebibliography\thebibliography
\def\thebibliography{\DeclareRobustCommand{\VAN}[3]{##3}\VANthebibliography}

\usepackage{graphicx}	
\usepackage{amsmath}	

\title[Modelling the CO streamers of Orion BN/KL]{Modelling the CO streamers in the explosive ejection of Orion BN/KL region}

\author[Rodr\'iguez-Gonz\'alez et al.]{A. Rodr\'iguez-Gonz\'alez$^{1}$,
\thanks{Contact e-mail: \href{mailto:ary@nucleares.unam.mx}{ary@nucleares.unam.mx}}, P.R. Rivera-Ortiz$^{2,3}$,
A. Castellanos-Ram\'irez$^4$,A.C. Raga$^1$,
\newauthor
L. Hern\'andez-Mart\'inez$^5$, J. Cant\'o$^4$,  L.A. Zapata$^3$\&  F. Robles-Valdez$^6$.
\\
$^1$Instituto de Ciencias Nucleares, Universidad Nacional Aut\'onoma de M\'exico, Ap. 70-543, 04510, Ciudad de M\'exico, M\'exico.\\
$^2$Univ. Grenoble Alpes, CNRS,
Institut de Plan\'etologie et d'Astrophysique de Grenoble (IPAG), 38000 Grenoble, France \\
$^3$ Instituto de Radioastronom\'ia y Astrof\'isica, Universidad Nacional Aut\'onoma de M\'exico,\\
Antigua Carretera a P\'atzcuaro 8701, Ex-Hda. San Jos\'e de la Huerta, Morelia, Michoac\'an, M\'exico C.P. 58089\\
$^4$ Instituto de Astronom\'ia, Universidad
Nacional Aut\'onoma de M\'exico, Ap. 70-264, 04510, Ciudad de M\'exico, M\'exico.\\
$^5$ Facultad de Ciencias.  Universidad Nacional Aut\'onoma de M\'exico, Ap. 70-399, 04510 Ciudad de M\'exico, M\'exico.\\
$^6$ Departamento de Astronom\'ia, Universidad de Guanajuato Ap. 144, 36000, Guanajuato, M\'exico.
}

\date{Accepted 13-Dec-22. Received 28-Sep-22 }

\pubyear{2015}

\begin{document}
\label{firstpage}
\pagerange{\pageref{firstpage}--\pageref{lastpage}}
\maketitle


\begin{abstract}
We present reactive gasdynamic, axisymmetric simulations of dense, high velocity
clumps for modelling the CO streamers observed in Orion BN/KL. We have considered 15 chemical species, a cooling function for atomic and molecular gas, and heating through cosmic rays. Our numerical simulations explore different ejection velocities, interstellar medium density configurations, and CO content. Using the CO density and temperature, we have calculated the CO ($J=2\to1$) emissivity, and have built CO maps and spatially resolved line profiles, allowing us to see the CO emitting regions of the streamers and to obtain position velocity diagrams to compare with observations. We find that in order to reproduce the images and line profiles of the BN/KL CO streamers and H$_2$ fingers, we need to have clumps that first travel within a dense cloud core, and then emerge into a lower-density environment.

\end{abstract}
\begin{keywords}
ISM: evolution-- ISM: kinematics and dynamics- ISM: molecules\end{keywords}
\section{Introduction}

Currently, the star formation paradigm considers that isolated pre-stellar cores evolve through accretion to become a star. In the case of low-mass stars, there is enough evidence to classify them from early to late protostars and their duration in the Class 0 through Class III objects. More massive forming stars seem to undergo a similar process, nevertheless, they are formed in dense environments which could produce close dynamical encounters that may even interrupt the accretion process. This effect has been recently invoked as the origin of explosive outflows, which have been reported to have occurred in at least three massive star-forming regions \citep{ZETAL11, ZETAL13, ZETAL17, ZETAL20}.  The closest of these explosive outflows is the Orion BN/KL region, located behind the Orion Nebula
\citep[at a distance of 414~pc from the Sun;][]{men2007}. The most accepted qualitative model for this object \citep{BETAL05} proposes that a single ejection event could have been
caused by a stellar merger or the dynamic rearrangement of a non-hierarchical system of
young, massive stars or protostars and, more quantitatively, now there is an active effort to explain the formation of these explosive outflows
\citep{TAN04,ZETAL09, GETAL11, BETAL11, RETAL21, ROETAL21}. 
\newpage
 The Orion BN/KL region has three dynamical components: a breaking-up multiple stellar systems,
an expanding molecular bubble, and a peculiar outflow with
about 200 filamentary structures, detected in H$_2$ and in CO ($J=2\to$1), known as ``fingers'' and ``streamers'', respectively. All components seem to have originated in the same event approximately 500~yr ago \citep{ZETAL09}. The fingers are the more extended and the first to be discovered. The Orion fingers were reported by \cite{AB93}  as H$_2$~2.1$\mu$m features
emanating outwards from the central region of Orion BN/KL and terminating in a series of Herbig-Haro (HH) objects,
which had previously been observed as [O~I]~6300, high-speed ``bullets'' by \cite{AT84}. These
HH objects have been detected in other optical lines \citep{ODELL97}. The
kinematics of the H$_2$ emission features has been studied with Fabry-Perot observations
\citep{CETAL97, SETAL99} and proper motion measurements \cite{BETAL11}.
The proper motions of some of the optically detected bullets have also been presented by
\cite{DETAL02}.

Following \citeauthor{BETAL11} (\citeyear{BETAL11} and \citeyear{BETAL15}) there are about 200 H$_2$ fingers, which present a distribution of longer features
to the NW, shorter fingers to the SE and SW, and very few and weak features to the E and NE.
The longer fingers have a length of $\sim 50000$~au (using the distance of 414~pc) and diameters
between 800 to 3200~au. The shorter filaments are narrower, more numerous, and tend to
overlap. The heads of the H$_2$ fingers (seen in H$_2$ and [Fe~II] IR lines and
optically as HH objects, see above) have diameters $\sim 40\to 400$~au.
The well-defined, longer filaments have velocities (derived from the radial
velocity and proper motion measurements) of $\sim 350$~km~s$^{-1}$.
The lengths and velocities of the fingers are consistent with an origin in a single
ejection event (for all fingers) $\sim 500\to 1000$~yr ago. However, there is evidence
of substantial braking of the motion of the heads of those fingers over their
evolution \citep{BETAL11}.  An estimation of the total kinetic
energy of the fingers is about $10^{47}\to 10^{48}$~erg, which
can be interpreted as an estimation of the energy of an ``ejection event'' that
gave rise to the present-day fingers \citep{BETAL11}.\\

This region also has an extended CO outflow which was first detected in single-dish observations by \citep{KS76}. In more recent interferometric observations, this outflow has been resolved into a system
of CO ``streamers'' \citep{ZETAL09, BETAL17}. The streamers show a more isotropic direction
distribution than the H$_2$ fingers, with several streamers travelling to the NW
emitting in the CO $J=2\to$1 rotational transition.
The total mass moving with velocities larger than 20 km~s$^{-1}$ is $\sim 8$~M$_\odot$ \citep[and references therein]{BETAL15}.
In order to propose a mean mass in each of the fingers, one can assume that all fingers have the same mass, therefore 
dividing total mass in this region by the $\sim 200$ observed fingers obtaining mass per finger of 
$\sim 0.04$~M$_\odot$ \citep{ROETAL19b}.\\

Many of the CO streamers partially coincide with the
H$_2$ fingers  but they do not reach out to the position of the optical ``bullets'' at the tip of
the fingers. Typically, the CO emission of the streamers fades away at a fraction $\epsilon=0.3\to 0.7$
of the length of the corresponding H$_2$ fingers. The CO streamers are barely resolved with the ALMA
interferometer, with widths of $400 \to 800$~au \citep{BETAL17}. Therefore, the CO streamers
are shorter and narrower than the H$_2$ fingers by a factor of $\sim 2$.
As some of the fingers and the streamers are spatially coincident and have different widths,
it can be argued that the CO emission is produced inside the H$_2$ fingers.\\

The radial velocities show a quite dramatic pattern of mostly red-shifted CO streamers to the W and SW,
blue shifted streamers to the N and E, and intermixed blue- and red-shifted streamers to the NW.
Also, the CO streamers have  ``Hubble law'', linear radial velocity vs. distance
signatures, which are also in good agreement with a simultaneous ejection $\sim 500$~yr ago \citep[see][]{ZETAL09}.
These ``Hubble law'' velocity vs. distance signatures indicate that the CO emitting material has
not suffered substantial braking. The peak radial velocities (at the tip of the CO streamers) have
values of less than $\sim 120$~km~s$^{-1}$, corresponding to approximately 1/2 of the fastest spatial motions
of the H$_2$ fingers (of $\sim 350$~km~s$^{-1}$, see above). However, it has been challenging to note that even when Orion H$_2$ fingers are larger than the CO streamers, they have different kinematic ages and do not follow a Hubble law. This could be explained by taking into account the interaction between the environment and the leading clumps that generate the fingers. \cite{ROETAL19b} used a dynamical model based on a plasmon equation of motion (\citealt{DYA67} and \citealt{ROETAL19}) to explain the age discrepancy and obtained an inferior limit for their ejection conditions.\\

The environment that surrounds the Orion BN/KL outflow is a dense molecular core. \citet{OETAL16} set a lower
density of $10^5$ cm$^{-3}$, analyzing the extinction and the overall angular spread of the H$_2$ emission.
From the CO emission, \cite{BETAL17} found an environment density between $10^6$ and $10^7$~cm$^{-3}$,
assuming a $X_{CO}=10^{-4}$ CO fraction and a background temperature of 10 K.
Lower values of $X_{CO}$ would lead to higher estimates of the environmental density.\\

The connection between the H$_2$ fingers and the CO streamers seems to be direct, both being created by leading bullets. Nevertheless, their differences need to be modelled and understood in order to explain and interpret the observations of similar explosive outflows, which are detected at high resolution with ALMA in the CO J=$2\to1$ line
but not in H$_2$ emission, such as DR21 and G5.89, which are at larger distances from us.\\

This paper presents simulations of the CO streamers in the Orion BN/KL region.
In our models, the streamers are produced by fast clumps travelling
within a structured environment. In particular, we study the effects
of having a central, ``dense cloud'' environment, from which the clumps
emerge into a lower-density medium. This configuration has been shown
to lead to the production of linear ramps in the predicted position-velocity
diagrams from the resulting flows \citep{RETAL22}. {\bf Notice that we deal with CO emission and compare the kinematic properties obtained from the numerical models with the observations. A larger study is necessary in order to better conciliate the CO and H2 emission, simultaneously.
}

The paper is organized as follows: in Section 2 we present the numerical setup of the simulations. In section 3, we present the results of the CO emission analysis. Finally, the contribution of these models to our understanding of the Orion BN/KL region is discussed in section 4.

\section{Numerical simulations}
We have computed 2D numerical simulations using {\sc Walkimya-2D}. The code solves the hydrodynamic equations (\citealt{Esquivel2009}) and a
chemical networks on an axisymmetric grid, The code is described in detail by \cite{CRETAL18} and \cite{ROETAL22}. 
The chemical network  tracks the abundances of 15 chemical species: non-equilibrium evolution of C, C$_2$, 
CH, CH$_2$, CO$_2$, HCO, H$_2$O, O, O$_2$, H$^+$, H$^-$, H and via conservation laws H$_2$, CO and OH are 
calculated. The selection of this network of species and the involved reactions is such that they can explain 
the CO abundances in other astrophysical situations (\citealt{CRETAL18}).

The complete set of equations for a reactive flow are presented in \cite{CRETAL18} and references 
therein. For a 2D flow, the reactive flow equations can be written as:
\begin{equation}
\frac{\partial \mathbf{U}}{\partial t}+\frac{\partial \mathbf{F}}{\partial
x}+\frac{\partial \mathbf{G}}{\partial y}=\mathbf{S},
\end{equation}
where $(x,y)$ are the (axial, radial) coordinates of the cylindrical domain, 
the vector $\mathbf{U}$ contains the so-called conservative variables,
$\mathbf{F}$ and  $\mathbf{G}$ are the fluxes in the $x$ and $y$-directions (respectively) and  $\mathbf{S}$ is
the sources vector. $\mathbf{S}$ contains the geometric terms resulting from the cylindrical grid
geometry, energy gain/loss terms (in the energy equation) and molecular formation/destruction rates (in
the continuity equations for the different chemical species).
We, therefore, calculate the reaction rates for each of the chemical species and
include  the thermal energy gain and loss due to interaction with the radiative field or
associated with the latent heat of the chemical reactions and/or the internal
energy of the molecular, atomic or ionic species.

The energy equation includes the cooling function described by \cite{RR04} for atomic gas and for lower temperatures 
we have included the parametric molecular cooling presented by \cite{KH07},

\begin{equation}
\label{eq:coolingmol}
\Lambda_{\rm mol}(T)=L_1 \cdot T^{\epsilon_1}+L_2 \cdot \exp \left ( -\frac{c_*}{(T-T_*)^{\epsilon_2}}\right) \;\;,
\end{equation}
for {{a temperature}} $T < 5280$~K, where, $L_1=4.4 \times 10^{-67}$, $L_2=4.89 \times 10^{-25}$, $c_*=3.18$, $\epsilon_1=10.73$, $\epsilon_2=0.1$ and 
$T_*=1.0$~K. The total radiative energy for temperatures lower than 5280~K is given by
$L_{rad,mol}=n_{\rm gas} \cdot n_{\rm CO} \cdot \Lambda_{\rm mol}(T)$, where $n_{\rm gas}$ and $n_{\rm CO}$ are the numerical densities of the 
gas and the CO molecule, respectively. The cosmic ray ionization rate of 
atomic hydrogen is  $\Gamma_{\rm crp}=5\times 10^{-28}n_{\rm H}$, see \cite{HETAL09}.

Finally, the thermal pressure is given by $P=(n+n_{e})kT\,,$ where $n$ is the total density of molecules+atoms+ions, $n_e$ is the electron density and {{and k is the Boltzmann constant}}.

\begin{figure}
    \centering
    \includegraphics[width=1.1\columnwidth]{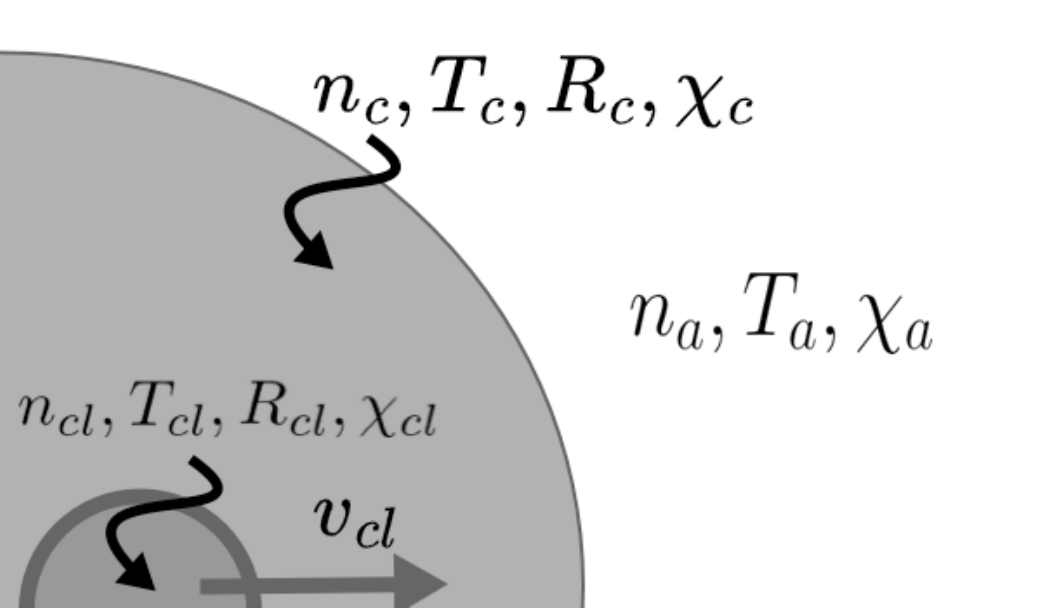}
    \caption{Schematic diagram showing the initial condition
      of the simulations. A fast clump initially travels within a dense cloud
      (of density $n_c$) and later emerges into a lower density, homogeneous
    environment (of density $n_a$).}
    \label{fig:model_scheme}
\end{figure}

{ \cite{BETAL15} modeled a ``CO streamer'' as the wake
  left behind by a fast, dense clump
  moving in a uniform environment. In the present paper, we study the
  case of a clump moving in an environment with a transition from
  an inner, dense region to an outer, lower-density region.
Our simulations follow the dynamics of a clump that is initially moving
  inside a dense core of density $n_{cl}$ and temperature $T_{cl}$ and then
  emerges into a homogeneous medium with number density n$_{a}$
  and temperature $T_{a}$. The CO compositions of the high velocity
  clump, the cloud core and the external environment
  and of the external environment are also
  considered as free parameters (see Figure \ref{fig:model_scheme}).}

The simulations are done on a binary adaptative mesh with 7 refinement
levels, yielding 
a maximum resolution of $4096 \times 1024$ (axial $\times$ radial) cells,
in a computational domain of
$48000 \times 12000$~au. Therefore, the maximum resolution of the
simulation is 11~au per pixel. 
We used reflective boundary conditions for the symmetry axis and 
a free outflow boundary condition for all the other frontiers.
The size of the mesh is large enough so that the choice of outer boundaries does not affect the simulation. 

For the initially spherical and homogeneous, fast clump, we choose:
\begin{itemize}
  \item a velocity $v_c=300\to 500$~km~s$^{-1}$,
  \item a radius $r_c=50$~au (in the range of sizes observed for the heads of the
    H$_2$ fingers),
  \item a mass $M_c=0.01$~M$_\odot$, resulting in an initial number density of H$_2$ molecules
    $n_c=10^{10}$~cm$^{-3}$,
  \item a temperature $T_c=30$~K.
\end{itemize}
These parameters are consistent with the H$_2$ and CO observations of the finger/stream system discussed above.

{Figure~\ref{fig:model_scheme} shows a schematic diagram of the components in the simulation: The initial fast clump of radius $R_{cl}$ moving at a velocity $v_{cl}$, with density $n_{cl}$, temperature $T_{cl}$, and molecular fraction $\chi_{cl}$. The clump is immersed in a static cloud of radius $R_{c}$, density $n_{c}$, at temperature $T_{c}$ and with a molecular fraction $\chi_{c}$. Finally, the density, temperature and molecular fraction of the external medium are $n_{a}$, $T_{a}$ and $\chi_{a}$, respectively.  }

In all runs, we assumed that the
chemical composition of the interstellar medium is a free parameter.
We have run 7 numerical simulations using different initial values of the density of the clump and initial CO compositions for each of the phases.

The fast clump travels along the $x$-axis, and
is initially located at 500~au from the left boundary of the computational
grid. It has an initial radius of 50~au, and has a $\rm CO$ abundance
 of
 1.67$\times 10^{-4}$ with respect to $\rm H_2$. The total gas density is given by adding the individual H$_2$, H, CO, C and O densities, or by adding the
 densities of atomic and molecular gas.  The clump has an initial temperature
of 30~K and a 0.03~M$_\odot$ mass. 

In all the models we have considered an external environment of density $n_a=10^5$~cm$^{-3}$ cm$^{-3}$
and temperature $T_a=1000$~K. We have included a central, dense core with a radius of
$R_{c}=2000$~au, density $n_{c}=100 \, n_a$ and in pressure equilibrium with the ambient medium.

The three components (high-velocity clump, central cloud, and outer
environment) are present in all models
except M0 and M00. In these two simulations, the clump
travels within a homogeneous environment of density $n_a$, and the dense,
central cloud is absent.
Table~\ref{tab:models} gives the molecular fraction of the three
initial components of the simulations.

 \begin{table}
\begin{center}
\caption{Chemical initial conditions of the numerical simulations} 
\label{tab:models}
{
 \begin{tabular}{l c c c l}
\hline \hline
\multicolumn{1}{c}{Model} &
\multicolumn{1}{c}{$\chi_a$} &
\multicolumn{1}{c}{$\chi_c$} &
\multicolumn{1}{c}{$\chi_{cl}$} \\

\hline

M00 & 0 & $--$ & 0.99\\
M0& 0.99 & $--$ & 0.99\\
M1& 0.99  & 0.99 & 0.99\\
M2& 0.99 & 0.99 & 0\\
M3& 0 & 0 & 0.99\\
M4& 0 & 0.99 & 0\\
M5& 0.1 & 0.99 & 0.1\\
\hline
\hline
\end{tabular}
}
\end{center}
\end{table}

\subsection{The CO emissivity}

We expect to have strong CO emission {arising}, mainly, from the three zones:
a) the high-velocity clump, b) the tail formed by the material left behind
by the clump, and c) the environmental gas shocked by the wings
of the bow shock around the clump. The simulations show that a substantial
amount of material is dragged out of the central, dense cloud (within
which begins the motion of the fast clump) as a result of the
passage of the clump bow shock. This process is fundamental in determining
the CO emission of the wake left behind by the fast clump.

To calculate CO emission maps and position-velocity diagrams
that can be compared with observations, we calculated the CO $J=2\to1$
emission coefficient in each of the
computational cells using the CO density and 
the temperature of the gas as 
\begin{equation}
 {j_{\rm {2 \to 1}}}=\frac{1}{4\pi}\frac{g_1}{g_2}n_{\rm CO} \cdot e^{\frac{-E_{\rm {2\to1}}}{k T_{{2\to1}}}} A_{\rm {2\to 1}} E_{\rm {2\to 1}},\\
\end{equation}
{\bf where, $g_1=3$ is a degeneracy factor, $Z(T)=\sum^N_i e^{-T_{\rm lev CO}/T(i,j)}$ is the partition function, with $T_{\rm lev CO}$ as the temperature of the corresponding energy levels,
  $A_{{2\to1}}=7.16 \times 10^{-7}$ s$^{-1}$ }
is the spontaneous emission coefficient,  $E_{{2\to1}}=h \nu_{{2\to1}}$ is the energy of the transition
with,  $\nu_{{2\to1}}=230.538$ GHz and $h$ is the
Planck constant \citep{Rybicki}.The emission maps and position-velocity diagrams
are obtained through appropriate integrals of the emission coefficient
along the lines of sight.

\subsection{CO maps and position-velocity diagrams}
Using the emission coefficients, we have constructed CO emission maps and
position-velocity (PV) diagrams to compare our numerical results with the observational data presented in the literature.

First, we have computed  CO ($J=2 \to 1$) emission maps by
appropriately rotating and integrating the axisymmetric simulations.
All of the CO maps discussed below
we have considered a $\phi=0$ orientation angle,
corresponding to an outflow axis on the plane of the sky.

We have also calculated spatially resolved line profiles, considering
the radial velocity and thermal Doppler profiles for the emission of
the computational cells. With these line profiles, we then computed
PV diagrams, integrating the emission perpendicularly to the projected
outflow axis. The PV diagrams discussed below have been computed
for different values of the orientation $\phi$ between the outflow
axis and the plane of the sky.

For the CO emission maps and PV diagrams, we have not included  the
CO emission of the gas with absolute velocities lower than 3 km/s.
This is to avoid contaminating our  {results} with the CO emission
coming from the unperturbed environment. This velocity cutoff
is similar to the one used by \citet{ZETAL09} for presenting
the CO emission of the Orion fingers.
\begin{figure}
\includegraphics[width=1.\columnwidth,trim={0 1.cm 1.5cm 2cm}, clip]{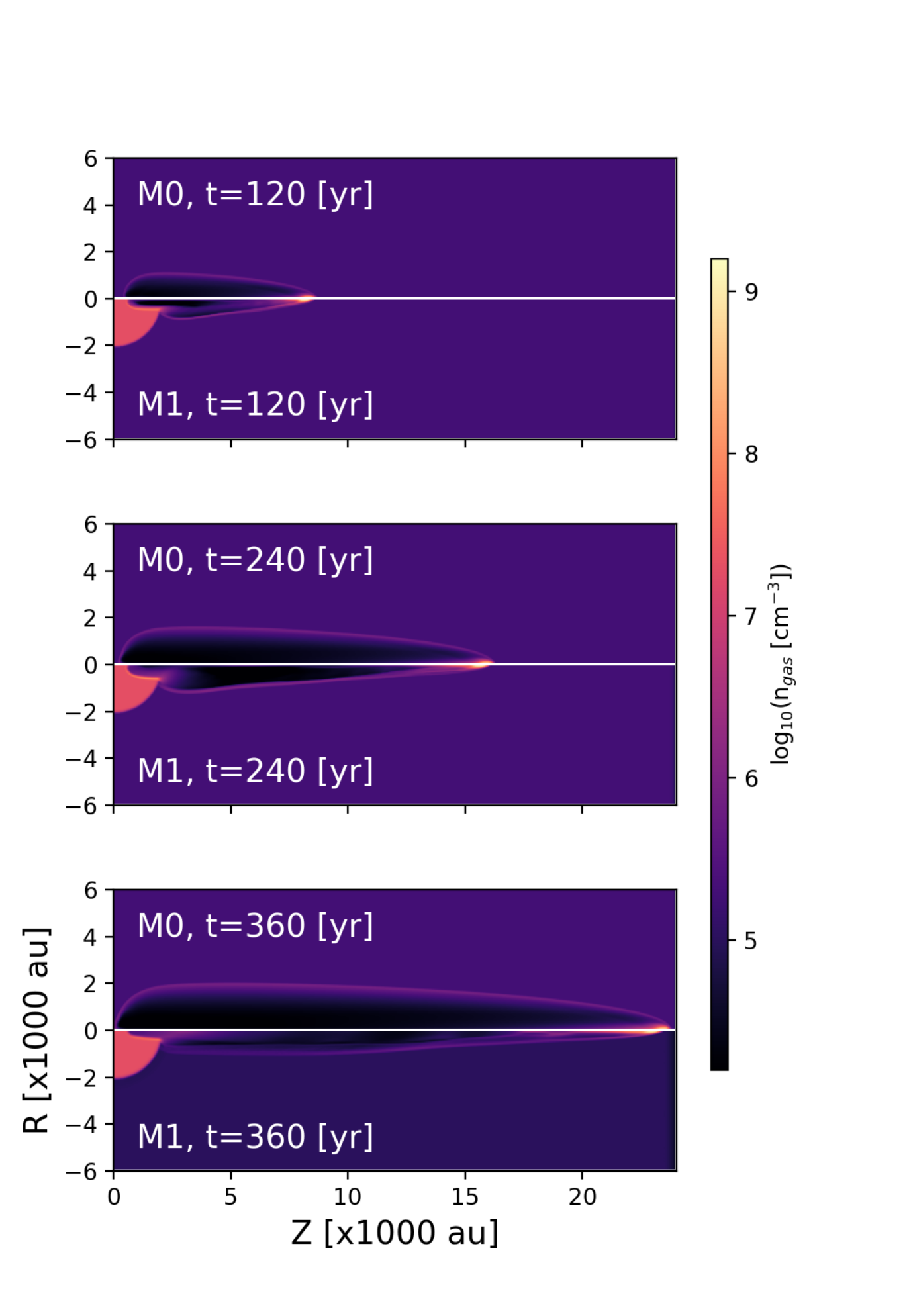}
\caption{Density stratifications obtained for models M0 (bottom half
  of each plot), and M1 (top half) at times 
  $t=$  $120$ (top), $240$ (centre) and $360$~yr (bottom). In model M1,
  the fast clump emerges from a central, dense cloud (which is not present
  in model M0). The logarithmic colour scale is shown by the bar on the right,
and the axial and radial axes are labelled in units of $10^3$ au.}
\label{fig:rhomaps}
\end{figure}

Finally, our emission maps and PV diagrams were 
convolved with a spatial gaussian profile with FWHM) corresponding
to the synthetic beamsize  of the ALMA interferometer of
$1\arcsec$ (corresponding to 400~au at the distance of Orion).

\section{Results}

We first discuss the results obtained for the M0 and M1 models, which
have the same initial chemical distribution (with molecular gas
in all of the initial computational domain). These two models differ
in that model M1 has a central dense cloud, and model M0 does not.

\begin{figure}
\centering
\includegraphics[width=1.1\columnwidth,trim={0 1.5cm 0 3cm}, clip]{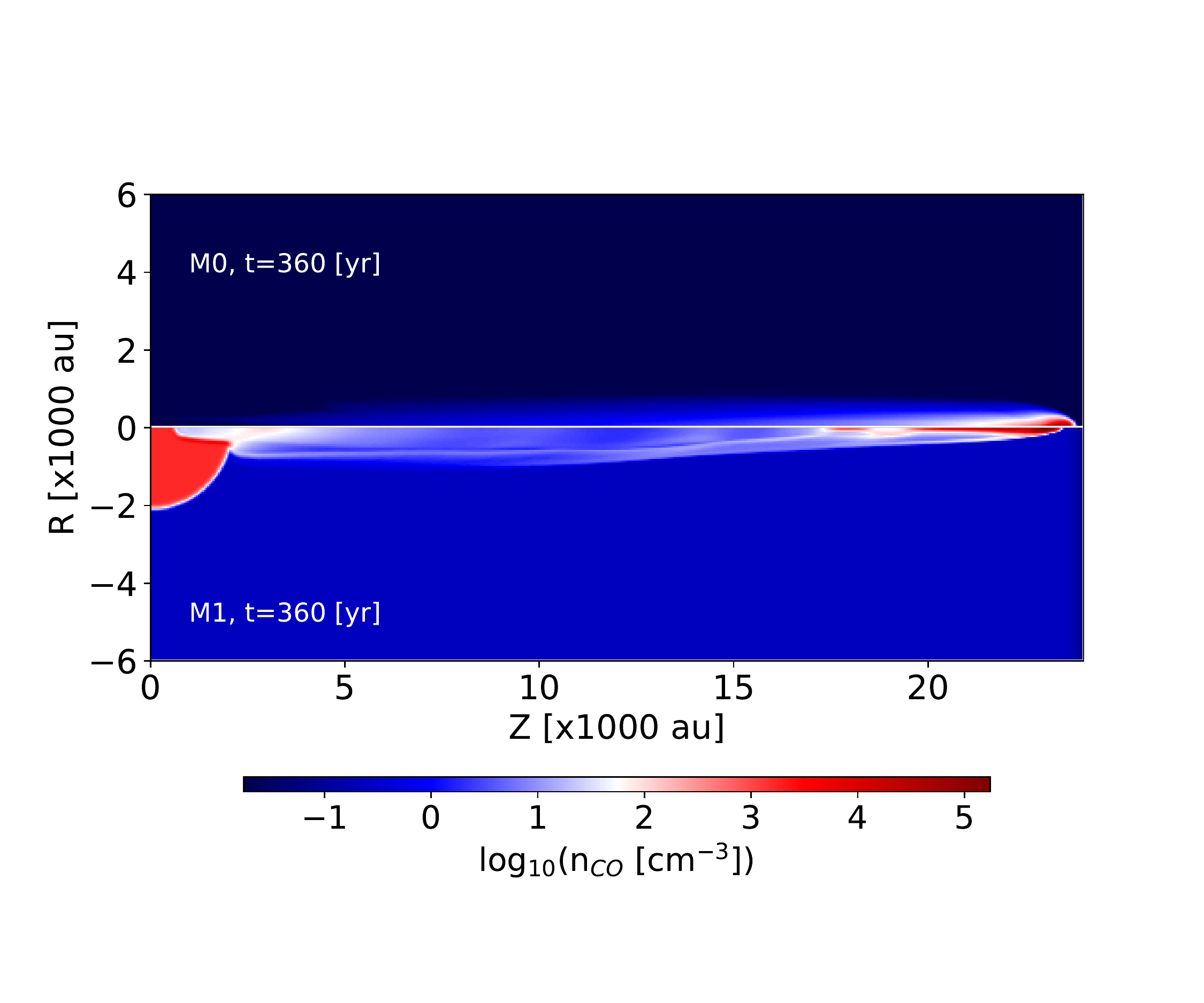}
\caption{CO density stratifications obtained from
Models M0 and M1 (upper and lower panel, respectively) 
for an evolutionary time of 360~yr. 
One can see that the highest density of CO is found in the clump, in the shocked ISM, in a small tail behind the clump and in a flow of material from the central cloud (for Model M1) that injects low-density material that produces the expansion of shock waves.}
\label{fig:map_co}
\end{figure}

\begin{figure}
\includegraphics[width=1.1\columnwidth,trim={0 1.5cm 0 3cm}, clip]{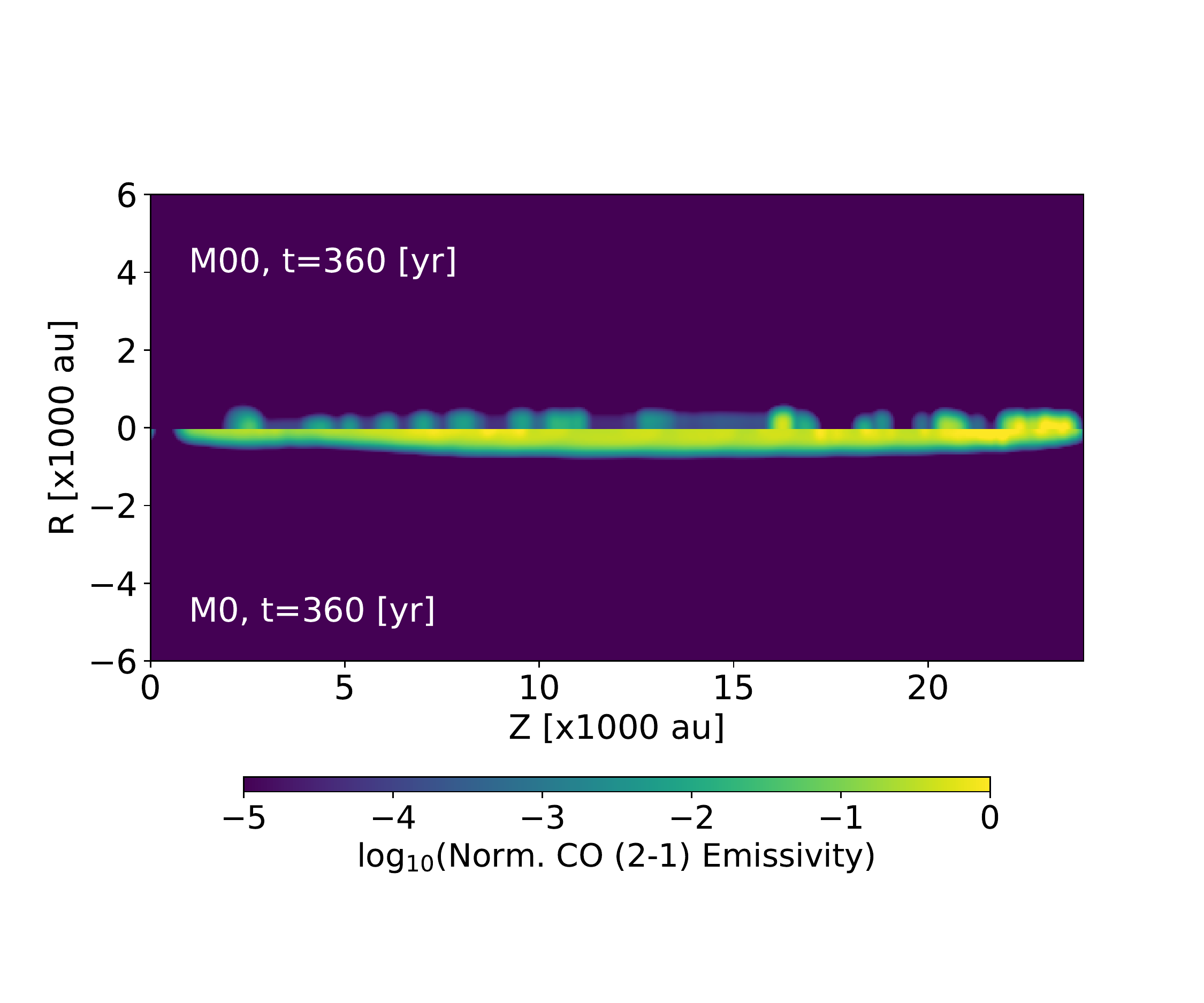}
\caption{The CO integrated emissivity, on the plane of the sky, for Models M0 and M00 (top and bottom panels, respectively), at an evolutionary time of 360 yr.}
\label{fig:emiss_m0m00}
\end{figure}
Figure~\ref{fig:rhomaps} shows the numerical density stratifications
of models M0 and M1 (upper and bottom panels, respectively) at 
evolutionary times of 120, 240, and 360~yr (top, middle and bottom panels, 
respectively). In all of the time frames, we see that the clump of
model M0 at all times:
\begin{itemize}
\item has a position that slightly trails behind the position of the
  M1 clump,
\item has a smaller radius,
\item has a radially more confined bow shock and wake region,
\end{itemize}
These are the three qualitative results of the passage through the
central, dense cloud present in the initial flow configuration in
model M1 (but absent in model M0).

\begin{figure}
\includegraphics[width=1\columnwidth]{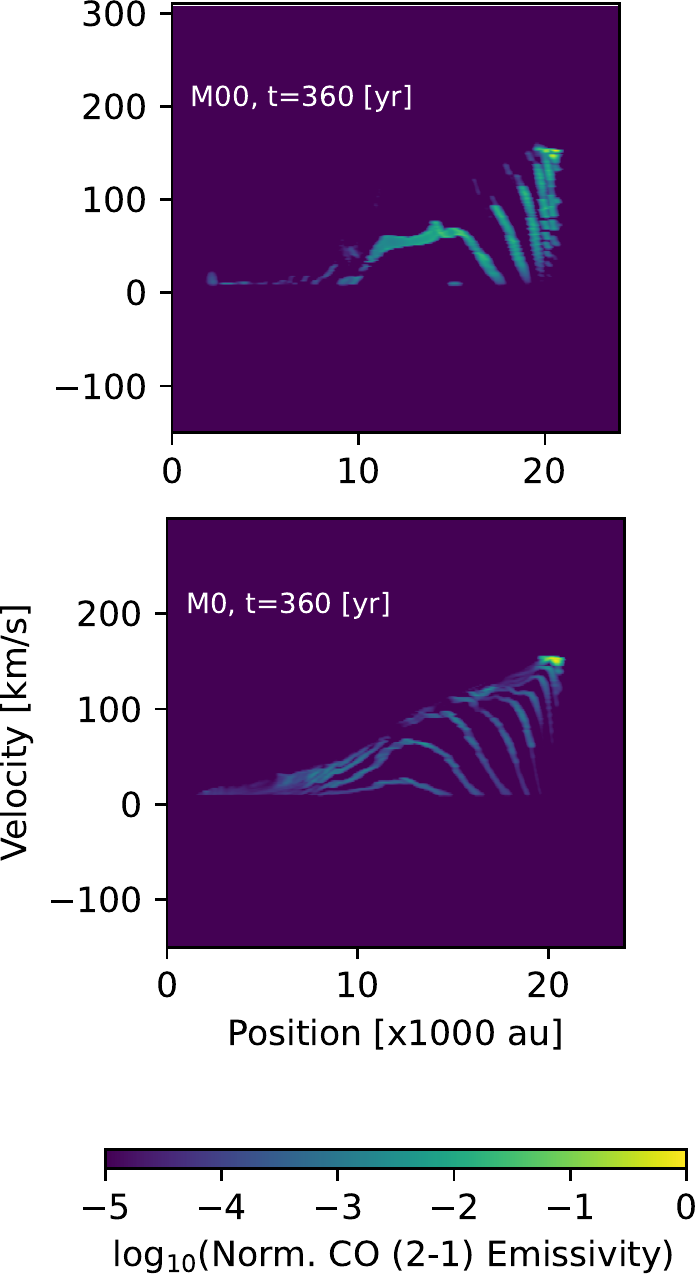}
\caption{Position-velocity (PV) diagrams obtained from models
  M00 (top) and M0 (bottom) for a $t=360$ yr evolutionary time
  assuming a $\phi$ = 30$^\circ$ angle between the outflow axis and the plane of the sky. The PV diagrams are normalized to the peak emission of each
  frame, and are shown with the logarithmic colour scale given by the bottom
  bar. The radial velocity is labeled in km~s$^{-1}$ and the axial coordinate
in $10^3$au}
\label{fig:pv_m0m00}
\end{figure}

Figure \ref{fig:map_co} shows the CO density stratification
for models M0 and M1 at t=360 yr.
the density structure is very different for these two models,
as expected. At M0 the highest molecular content  is found mainly
{\bf in the leading clump}, that was originally blown, and a CO plume is visible (mainly) in the shocked gas region. And for model M1, it can be observed (as well as gas density) a stream of CO coming out from the central cloud inserted in the empty region behind the clump, and the fingertip does not have the maximum CO density of the model. This maximum is found in the central cloud and in the steam coming from there because of the pressure gradient between the cloud and the internal region of the bow shock that produces a vacuum.


Figure \ref{fig:emiss_m0m00} shows CO ($J = 2\to1$) emission maps
obtained from models M00 and M0 (top and bottom
panels, respectively) at a $t=360$~yr
evolutionary time, assuming that the outflow axis lies on the
plane of the sky. For model M00 (which has a fast, molecular clump
moving in a uniform atomic environment, see Table~\ref{tab:models}), the CO emission
comes from the fast clump, and from detrained clump material which has
been left behind in the wake of the clump, {\bf that is, material that has been entrained from the central core by the clump, that eventually slows down, filling the wake}. For model M0 (with a molecular
clump moving in a uniform molecular environment, see Table~\ref{tab:models}) the
CO emission also has a strong contribution from environmental
material shocked by the wings of the bow shock around the fast clump.

In Figure \ref{fig:pv_m0m00}, we present the PV diagrams (showing the
line profiles integrated across the outflow axis, as a function of position
along this axis) obtained from models M00 and M0 (top and bottom
panels, respectively), assuming
an angle of $30^\circ$ between the outflow direction and the
plane of the sky, for a $t=360$~yr evolutionary time.

In these PV diagrams, the peak emission comes from the
fast clump itself. The region between the clump and the outflow
source is occupied by a series of arcs (in the PV plane), produced
by the detrained clump material in the wake of the clump. The
emission within this region has an upper envelope of increasing
peak radial velocity as one moves away from the outflow source.
These models do not produce the  linear velocity vs. position ramp
observed in the Orion fingers (see, e.g., \citet{ZETAL09}).

\begin{figure}
\includegraphics[width=1.1\columnwidth,trim={0 1.5cm 0 3cm}, clip]{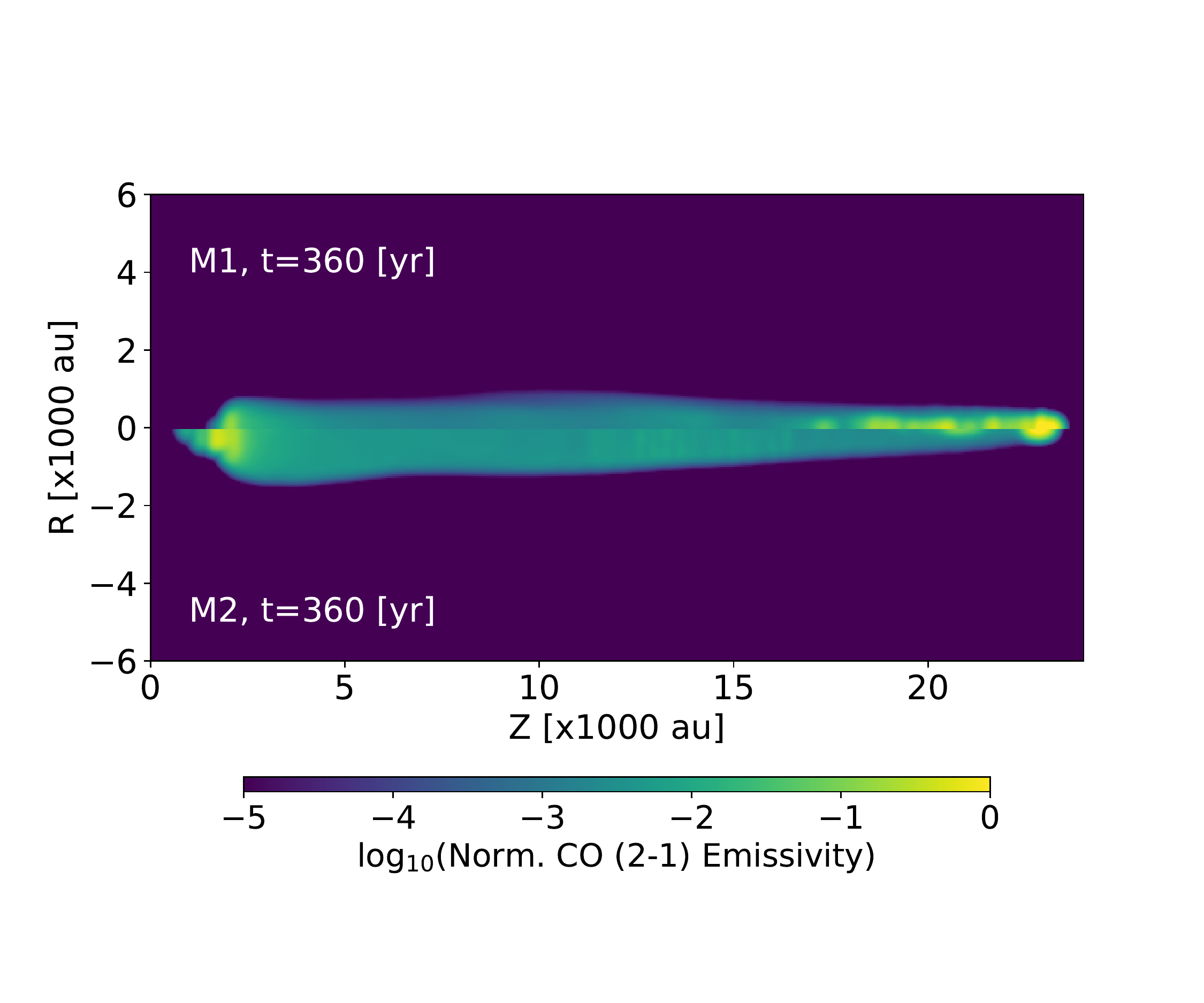}
\caption{The CO for Models M1 and M2 (top and bottom half, respectively),
  at a $t=360$~yr evolutionary time, computed assuming that the outflow
  axis lies on the plane of the sky. The emission (shown with the logarithmic
  colour scale given by the bottom bar) is normalized to the peak intensity
  in each of the maps. The plane of the sky coordinates is in units
of $10^3$au.}
\label{fig:emiss_m1m2}
\end{figure}

We now present CO $J=2-1$ intensity maps and PV diagrams from models
M2-M5, in all of which the fast clump first travels through a central,
dense cloud and then emerges into a lower-density, uniform environment.
Figure \ref{fig:emiss_m1m2} shows the CO emission maps
(assuming that the outflow axis is on the plane of the sky)
for models M1 and M2, at $t= 360$ yr (top and bottom panels, respectively).
In these models (unlike the models that do not have a central cloud,
see Figure \ref{fig:emiss_m0m00}), the CO emission fills the cavity behind the clump
bow shock, and has the peak emission in the region in which the
flow emerges from the dense cloud. In model M1 (in which all of the
initial components are molecular, see Table~\ref{tab:models}), a peak in the CO
emission is also observed at the position of the fast clump. This emission
peak is absent in model M2, which has an initially neutral fast
clump (see Table~\ref{tab:models} and Figure~\ref{fig:emiss_m1m2}).

\begin{figure}
\includegraphics[width=1.\columnwidth]{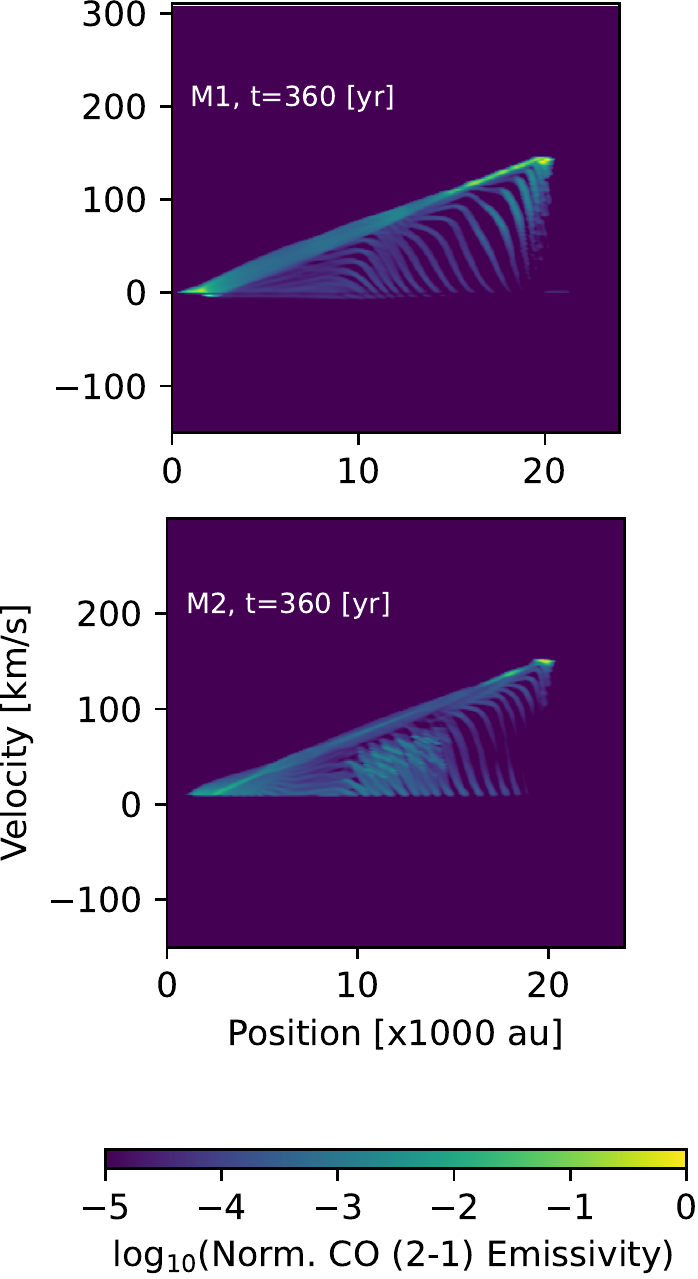}
\caption{PV diagrams obtained from models M1 and M2 (top and bottom panel, respectively) at $t=360$~yr, assuming a $\phi$ = 30$^\circ$ angle between the outflow axis and the plane of the sky.}
\label{fig:pv_m1m2}
\end{figure}

It is clear that in models M1 and M2 we obtain CO PV diagrams with a linear
ramp of emission with increasing radial velocities as a function of
position (see Figure \ref{fig:pv_m1m2}, beginning at the outer edge of the dense, central
cloud, and ending at the position-velocity position of the
high-velocity clump. This ramp of emitting material comes from a mixture
of detrained clump material and dense cloud material perturbed by the
passage of the clump.

\begin{figure}
\includegraphics[width=1.1\columnwidth,trim={0 1.5cm 0 3cm}, clip]{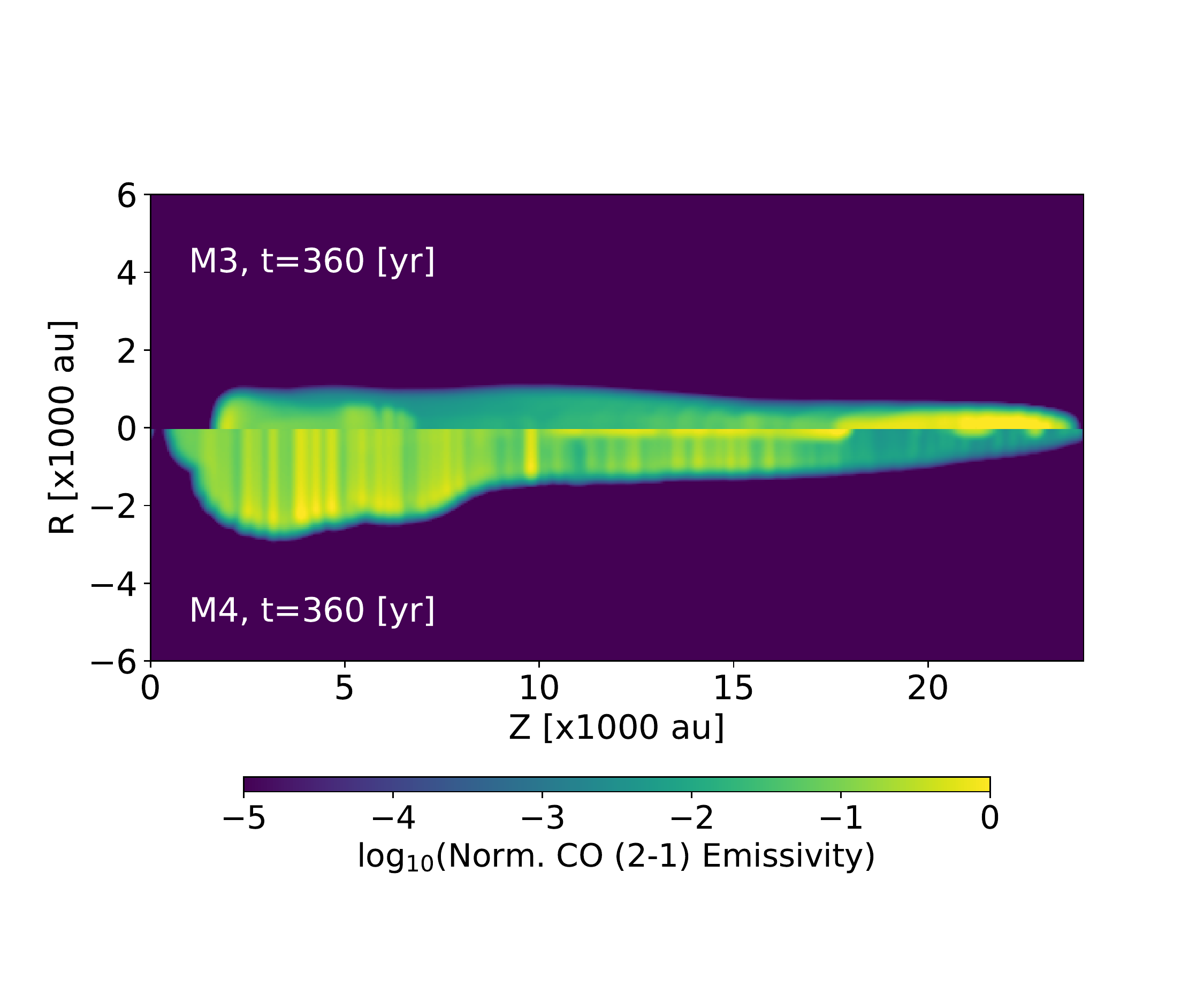}
\caption{CO emission maps from Models M3 and M4 (top and bottom panels, respectively) at a $t=360$~yr evolutionary time, assuming that the outflow axis
lies on the plane of the sky.}
\label{fig:emiss_m3m4}
\end{figure}

\begin{figure}
\includegraphics[width=1.\columnwidth]{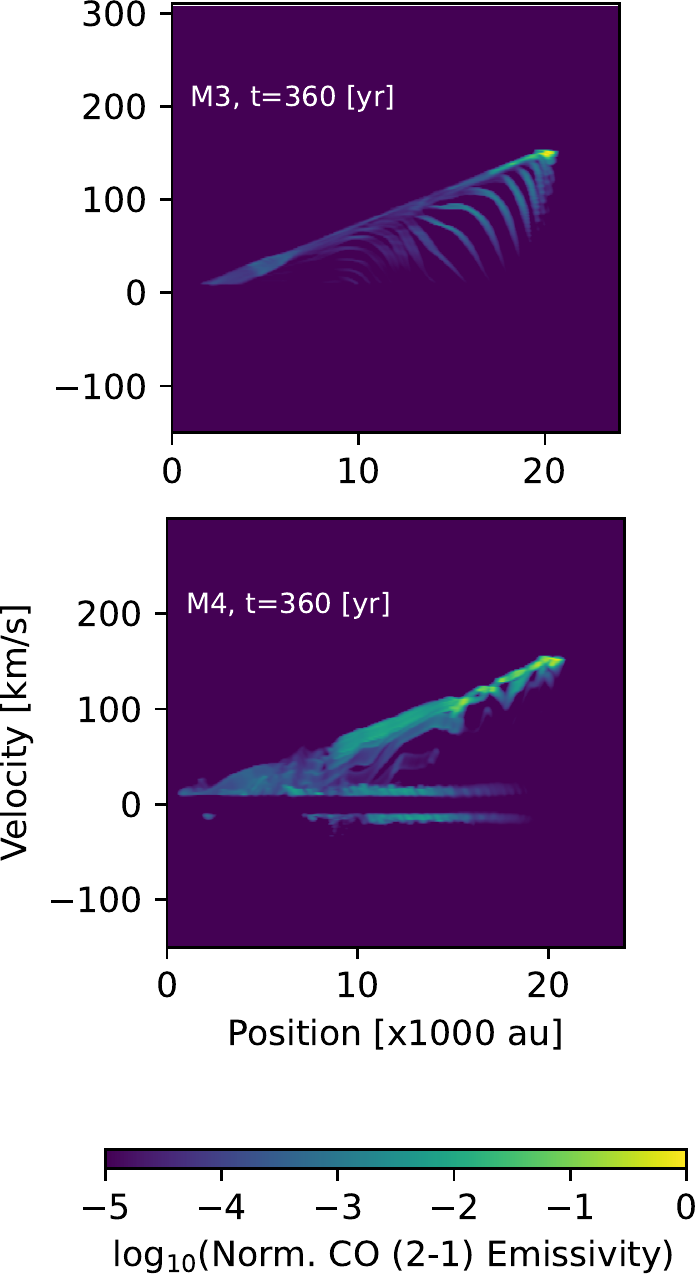}
\caption{PV diagrams obtained for models M3 and M4 (top and bottom panel, respectively) at $t=360$~yr, assuming a $\phi$ = 30$^\circ$ angle between the outflow axis and the plane of the sky.}
\label{fig:pv_m3m4}
\end{figure}

\begin{figure}
\includegraphics[width=1.\columnwidth]{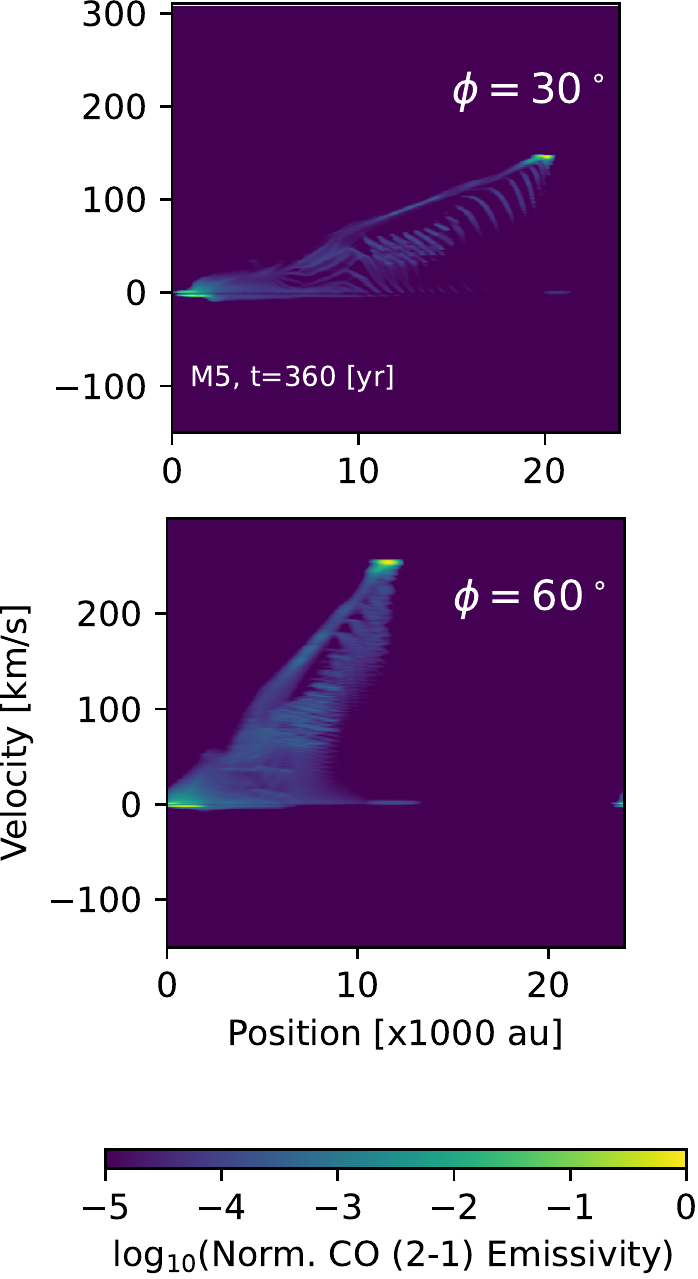}
\caption{PV diagrams obtained from model M5 for $t=360$~yr, assuming
  $\phi$ = 30$^\circ$ (top) and 60$^\circ$ angle (bottom) between the outflow axis and the plane of the sky.}
\label{fig:pv_m5}
\end{figure}

Figure \ref{fig:emiss_m3m4} shows a clear difference between the CO emission maps
(at $t=360$~yr) of models M3 (with neutral cloud/environmental gas) and
M4 (with molecular central cloud and neutral outer environment). The
PV diagrams obtained from these two models (see Figure \ref{fig:pv_m3m4}) also show
a linear velocity vs. position emission ramp (see Figure \ref{fig:pv_m5}). Finally,
model M5 (with partially molecular central cloud and fast cloud
and fully molecular outer environment) also produces PV diagrams
with an approximately linear velocity vs. position ramp (see Figure \ref{fig:pv_m5}).

From this exercise, we conclude that the presence of an inner, dense
cloud in the environment (through which is travelling the fast clump)
results in PV diagrams with an approximately linear ramp of
increasing velocities as a function of distance from the outflow
source. This feature is quite persistent, as it is present even in
the case in which only the material of the fast clump is molecular
(model M3).

{\bf Our results are clearly dependent on the density structure. Morphologically, filaments are formed in models from M0 to M5, but the kinematic differences point towards the structure present in models from M1 to M4. This is the behaviour shown by the observational PV diagrams, with a linear dependence of the velocity with the position (Zapata et al. 2009, figure 2, and Bally et al. 2017, figure 5).}

\section{{\bf Summary}}

Explosive outflows seem to be the result of a poorly understood
interaction of  protostellar objects in massive star-forming regions.
To interpret the observed characteristics of the explosive outflow
Orion BN/KL we present a set of simulations of the "CO streamers",
assuming an ejection of high-velocity clumps which interact with
a structured ambient medium.  
We carried out reactive flow numerical simulations solving the rate
equations of  15 species (H, H$^+$, C, C$_2$, CH, CH$_2$, CO$_2$,
HCO, H$_2$O, O, O$_2$, H$_2$, CO, OH, and e$^{-}$ ) considering 47
chemical reactions \citep[as in][]{CRETAL18}, and including
a molecular/atomic/ionic cooling rate.

In this work, we have focused on studying the importance of:
\begin{itemize}
\item a) the possible presence of a dense, central cloud within which
  the high-velocity ejecta starts to evolve,
\item b) the effect of the initial molecular fractions in each of the
  components: ejected clump, central cloud and outer, uniform
  interstellar medium.
\end{itemize}

In order to compare our numerical results with the observational features
of the Orion fingers presented in \citet{ZETAL09} and \cite{BETAL17},
we have obtained predictions of CO ($J=2\to$1) intensity maps and PV
diagrams. To this effect, we have calculated the appropriate emission
coefficient (from the temperature and non-equilibrium CO abundance
in each computational cell) and carried out the appropriate line of
sight integration. Clearly, this process could be also done for
many other molecular lines, but our present paper is restricted
to the CO ($J=2\to$1) line.


Our models of fast clumps travelling in a uniform-density medium
(models M00 and M0, see Table~\ref{tab:models}) produce emission maps with CO
streamers with widths of $\sim 1000$~au (see Figure \ref{fig:emiss_m0m00}). The PV
diagrams predicted from these models show the fast velocity
clump, and a highly structured wake joining the clump to the outflow
source.

Our models with a central ``dense cloud'' region produce emission maps
with CO streamers of varying widths (see Figures \ref{fig:emiss_m1m2} and \ref{fig:emiss_m3m4}), with the
narrowest streamers for the case of model M3, in which only the
fast clump is molecular (see the top panel of Figure \ref{fig:emiss_m3m4} and Table~\ref{tab:models}).
All of these models produce CO PV diagrams with a clear, linear
velocity vs. position ramp joining the edge of the central, dense cloud
to the PV position of the fast clump. This ``Hubble law'' expansion
is a quite sturdy feature, because it is present in cases in which
the central cloud is molecular (models M1, M2 and M4) or not
(model M3, in which only the fast clump is molecular, see Table~\ref{tab:models}).
This linear ramp in the PV diagrams is in clear qualitative agreement
with the structures observed in the Orion fingers.

It is important to stress that the calculated PV diagrams
show an expansion ramp that does not point to the origin of the clump
motion, but to the outer edge of the central, dense cloud. This result
is also in qualitative agreement with the observations of the CO streamers,
in which each of PV ramps of the fingers points to a central volume
of a few thousand astronomical unities as reported by \citet{ZETAL09, ZETAL20}. This
possibly corresponds to a central cloud from which have emerged
the clumps that formed the fingers in the Orion BN/KL region.

\section*{Acknowledgements}
We acknowledge support of the UNAM-PAPIIT grants IN110722, IN103921, IN113119, IG100422 CONACYT grant 280775 and also the Miztli-UNAM supercomputer project LANCAD-UNAM-DGTIC-123 2022-1. A. Castellanos-Ram\'irez and F. Robles-Valdez acknowledge support from CONACyT postdoctoral fellowship. P. R. Rivera-Ortiz acknowledges support from DGAPA-UNAM postdoctoral fellowship.

\section{DATA AVAILABILITY}
The data underlying this article will be shared on reasonable request to the corresponding author.

\end{document}